\documentclass[onecolumn,appendixfloats]{aastex62}
\usepackage{amsmath}

\begin{document}


\title{Fragmentation of Filamentary Cloud Permeated by Perpendicular Magnetic Field II. \\
Dependence on the Initial Density Profile}

%% Use \author, \affil, plus the \and command to format author and affiliation 
%% information.  If done correctly the peer review system will be able to
%% automatically put the author and affiliation information from the manuscript
%% and save the corresponding author the trouble of entering it by hand.
%%
%% The \affil should be used to document primary affiliations and the
%% \altaffil should be used for secondary affiliations, titles, or email.

%% Authors with the same affiliation can be grouped in a single
\author[0000-0002-7538-581X]{Tomoyuki Hanawa}
\affil{Center for Frontier Science, Chiba University,
1-33 Yayoi-cho, Inage-ku, Chiba,
Chiba  263-8522, Japan}

\author[0000-0003-1292-940X]{Takahiro Kudoh}
\affiliation{Faculty of Education, Nagasaki University, 1-14 Bunkyo-machi, Nagasaki,
Nagasaki 852-8521, Japan}

%% Use the \and command so offset the last author.
%\and

\author[0000-0003-2726-0892]{Kohji Tomisaka}
\affiliation{Division of Theoretical Astronomy, National Astronomical Observatory of Japan,
2-21-1 Osawa, Mitaka, Tokyo 181-8588, Japan}
\affiliation{Department of Astronomical Science, School of Physical
Sciences, SOKENDAI (The Graduate University for Advanced Studies), Mitaka, Tokyo 181-8588, Japan }

\begin{abstract}
We examine the linear stability of a filamentary cloud permeated by a perpendicular magnetic field.
The initial magnetic field is assumed to be uniform and perpendicular to the
cloud axis.   The model cloud is assumed to have a Plummer-like density profile and to be supported against the self-gravity
by turbulence.   The effects of turbulence are taken into account by enhancing the effective pressure of a low density gas.  
We derive the effective pressure as a function of the density from the condition of the
hydrostatic balance.  It is shown that the model
cloud is more unstable against radial collapse, when the radial density slope is shallower.
When the magnetic field is mildly strong, the radial
collapse is suppressed.   If the displacement vanishes in the region very far from
the cloud axis, the model cloud is stabilized completely by a mildly strong magnetic field.  
If rearrangement of the magnetic flux tubes is permitted, the model cloud is unstable even
when the magnetic field is extremely strong.  The stability depends
on the outer boundary condition as in case of the isothermal cloud. 
The growth rate of the rearrangement mode is smaller when the radial density slope is shallower.
\end{abstract}




\section{INTRODUCTION} \label{sec:intro}

Filamentary structures are ubiquitously found in the star-forming regions 
\citep[see, e.g.,][and the references therein]{andre14}.   They are considered
as an intermediate state from clouds to stars and the fragmentation is likely
to be a process forming cores.  This idea is supported
by observations showing that  prestellar cores and newly formed stars are associated with
the dense parts of filamentary clouds.   Although filamentary clouds are unstable
against fragmentation in general \citep[see, e.g.,][]{stodolkiewicz63,larson03},
a magnetic field may suppress the fragmentation if it is strong and perpendicular
to the cloud axis.   The effect of the magnetic field on fragmentation is a key
issue for understanding the core formation.

We have examined the stability of a filamentary cloud permeated by a perpendicular
magnetic field  against fragmentation using a simplified model
\citep[paper I in the following]{hanawa17}.  The initial magnetic field was assumed to
be uniform and the gas was assumed to be isothermal in paper I for simplicity.
These assumptions are made from technical reasons that it is difficult to make
an equilibrium model for molecular cloud permeated by a perpendicular cloud
 \citep[see, e.g.,][]{tomisaka14,hanawa15}.
When the magnetic field is parallel to the cloud axis or helical around the axis, 
we can make various equilibrium models by assuming symmetry around the cloud axis
\cite[see, e.g.,][]{toci15b}.  
Such model clouds have been studied extensively for many years 
\citep{stodolkiewicz63,nakamura93,hanawa93,fiege00}. However, the magnetic
fields are perpendicular to denser clouds \citep[see, e.g.,][]{sugitani11,andre14,kusune16,soler16}, 
although less dense clouds are associated with parallel magnetic fields.

Magnetic field direction is important to fragmentation of a filamentary cloud.
A magnetic force is perpendicular to the magnetic fields and hence gas flow along the axis
can not be suppressed by magnetic fields parallel to the axis.   
The wavelength of fragmentation is shorter when the parallel magnetic
field is stronger.  This apparent destabilizing effect is ascribed to the fact that the
magnetic field is assumed to be concentrated around the axis to support the
cloud against the radial collapse.  Given the central density and temperature, the filament 
diameter is larger for a stronger magnetic field.  The wavelength of the fragmentation, which
is roughly twice of the Jeans length, is shorter for a stronger magnetic field, when
measure in unit of the diameter. 
Moreover, the mass to flux
ratio is infinitely large, and hence  supercritical  if the cloud is elongated along the magnetic
field.  When the magnetic field is perpendicular to the cloud axis, the magnetic force
works against fragmentation (paper I).   Even mildly strong magnetic fields
stabilize a filamentary cloud against fragmentation, if they are perpendicular to the cloud
and their ends are fixed in the region very far from the cloud.   When the magnetic field
is helical around the axis, the magnetic force works against fragmentation but induces
non-axisymmetric instability \citep[see, e.g.,][]{hanawa93,fiege00}.

Interestingly perpendicular magnetic fields suppress instability less effectively
if the field lines are free, i.e., allowed to move.   When the magnetic fields
are free in the region very far from the cloud, they are rearranged to fragment the cloud
by reducing the gravitational energy of the cloud.  This means that the stability depends
on the outer boundary condition. 
When the magnetic fields are parallel to the cloud axis, boundary condition has little
effects on the instability unless they are placed close to the cloud axis.
This is reasonable since the instability is due to the self-gravity of the cloud and depends only
on the dense central gas.

In paper I  the model cloud is assumed to be isothermal and supported by gas pressure alone 
against gravity in equilibrium.  Accordingly, the density is assumed to decrease in proportion
to $ r ^{-4} $ in the region very far from the cloud center, where $ r $ denotes the distance from the cloud axis.
However, observed clouds show much shallower radial density profiles, which is often approximated
by a Plummer-like one,
\begin{eqnarray}
\rho (r) & = & \rho _c \left[ 1 + \left( \frac{r}{R _{\rm flat}} \right) ^2 \right] ^{-p/2} ,  \label{plummer11}
\end{eqnarray}
where $ \rho _c $ and $ R _{\rm flat} $ denote the central density and \lq radius\rq, respectively 
\citep{arzoumanian11,juvela12,palmeirim13,ohashi18}.
The index, $ p $, denotes the slope of the density profile, $ - d\ln \rho/d\ln r $, in the region
far from the cloud axis.   The index is estimated to be $ p \approx 2 $ from the model fit to
the column density distribution derived from the sub-millimeter continuum emission.

Considering the above mentioned arguments, we examine the stability of the Plummer-like cloud
against fragmentation taking account of perpendicular magnetic fields.
We assume that the Plummer-like profile is supported against
gravity by \lq effective gas pressure\rq, which mimics effects of turbulence.    If the effective 
temperature decreases with increase in the density, the Plummer-like profile with $ p < 4 $ is
realized as will be shown later.  We examine the effects of the density profile on fragmentation
of a filamentary cloud using the method developed in paper I.   When $ p < 4 $, the model
cloud is unstable also against radial collapse, although the isothermal model is neutrally
stable against it \citep[see, e.g., the review by][]{larson03}.   
It will be shown that the radial collapse is stabilized by mildly strong magnetic
fields.   The growth rate of the instability depends on the index, $ p $, but only quantitatively
except for the radial collapse.   

This paper is organized as follows.  We describe our assumptions and methods for computation
in \S 2.   The results are shown in \S 3, where also the stability of unmagnetized cloud is analyzed.
We discuss the implications of our models in \S 4 and summarize our main findings
in \S 5.  Appendices \ref{BesselK} and \ref{1Dmode} are
devoted to improvement for computing long wavelength modes and computing the case of
no magnetic fields, respectively.

\section{METHODS}

\subsection{Basic Equations}

As in paper I, we employ the ideal magnetohydrodymamic (MHD) equations for
our stability analysis.  They are expressed as
\begin{eqnarray} 
\frac{\partial \rho}{\partial t} + \mbox{\boldmath$\nabla$} \cdot
\left( \rho \mbox{\boldmath$v$} \right) = 0 ,  \label{continuity} \\
\rho \frac{d \mbox{\boldmath$v$}}{dt} = - \mbox{\boldmath$\nabla$} P 
+ \mbox{\boldmath$j$} \times \mbox{\boldmath$B$} -
\rho \mbox{\boldmath$\nabla$} \Phi ,  \label{motion_eq} \\
\mbox{\boldmath$j$} = \frac{\mbox{\boldmath$\nabla$} \times \mbox{\boldmath$B$}}{4 \pi} , 
\label{e-current}
\end{eqnarray}
where $ \rho $, $ \Phi $, $ \mbox{\boldmath$v$}$, $ \mbox{\boldmath$B$} $, 
and $ \mbox{\boldmath$j$} $ denote the density, gravitational potential, velocity,
magnetic field, and electric current density, respectively.    Here the symbol, $ P $,
denotes the pressure, which is designed to include effects of turbulence implicitly.

We ignore ambipolar diffusion for simplicity.  The ambipolar diffusion weakens the
magnetic force \cite[see, e.g.][]{hosseinirad18}.  However, the typical timescale
is a factor ten longer than the dynamical timescale \cite[see, e.g.,][]{nakano88} and
the effects are not large.  Thus we do not take account of the ambipolar diffusion
in order to avoid further complication.  Note that the ambipolar diffusion does not
work in our initial model since the magnetic field is uniform.  
The ambipolar diffusion works works only on a perturbation.  Thus, it
reduces the growth rate of instability but cannot suppress the stability.   See e.g.,
\cite{hosseinirad18} for the effects of ambipolar diffusion on the fragmentation.
They analyzed the stability of a filamentary cloud permeated by longitudinal
magnetic field.

\begin{figure}
\epsscale{0.5}
\plotone{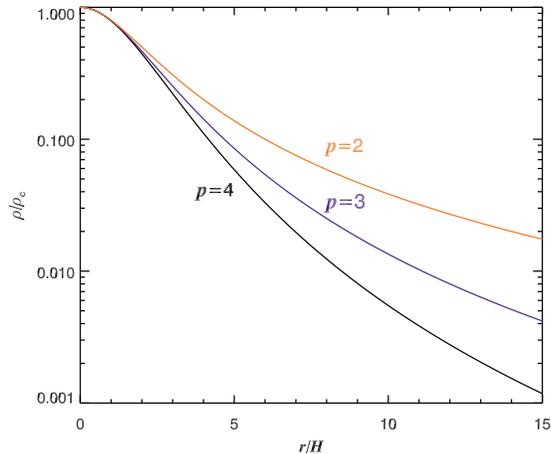}
\caption{Equilibrium density profiles for the index of $ p $ = 2, 3, and 4. \label{r-density}}
\end{figure}

The equation of state is specified in the subsequent section so that the
equilibrium density distribution is well approximated by the Plummer function,
\begin{eqnarray}
\rho _0 & = & \rho _c \left( 1 + \frac{r ^2}{2 p H ^2} \right) ^{-p/2} , \label{plummer}
\end{eqnarray}
where $ \rho  _c $ and $ r $ denote the central density of the filamentary cloud and 
the distance from the cloud axis, respectively.   The symbols, $ p $ and $ H $, denote
the index and length scale, respectively.  The length scale given in Equation (\ref{plummer11})
is expressed as $ R _{\rm flat} = \sqrt{2p} H $.  The radial density profile is shown in
Figure~\ref{r-density}.

The self-gravity of the gas is taken into account 
through Poisson's equation, 
\begin{eqnarray}
\Delta \Phi = 4 \pi G \rho ,
\end{eqnarray}
where $ G $ denotes the gravitational constant.  

In the following, we use the unit system where $ \rho _c = 1 $, $ H = 1 $ and $ 4 \pi G \rho _c  = 1 $,
in our numerical computations.

\subsection{Equilibrium Model}

When the density profile is expressed by Equation (\ref{plummer}), the filamentary cloud has
the mass per unit length,
\begin{eqnarray}
\lambda _r & = & 2 \pi \int _0 ^r \rho _0 \left( r ^\prime \right) r ^\prime dr ^\prime \\
& = & 
\begin{cases}
\displaystyle 4 \pi \rho _c  H ^2  \ln \left( 1 + \frac{r ^2}{4 H ^2} \right)   & (p = 2) \\
\displaystyle \frac{4 \pi p \rho _c H ^2}{p - 2}  \left[ 1 - \left( 1 + \frac{r ^2}{2 p H ^2} \right) ^{1-p/2} \right] 
& \mbox{(otherwise)}
\end{cases} ,  \label{linemass}
\end{eqnarray}
inside the radius, $ r $.  Thus the gravity is evaluated to be
\begin{eqnarray}
g _r & = & - \frac{d\Phi}{dr} = - \frac{2 G \lambda _r}{r} \\
& = & 
\begin{cases}
- \displaystyle \frac{8 \pi G \rho _c H^2}{r} \ln  \left( 1 + \frac{r ^2}{4 H ^2} \right)   & (p = 2) \\
- \displaystyle \frac{8 \pi G p \rho _c H ^2}{\left(p - 2 \right) r}  \left[ 1 - \left( 1 + \frac{r ^2}{2 p H ^2} \right) ^{1-p/2} \right] 
& \mbox{(otherwise)}
\end{cases} . \label{gravity}
\end{eqnarray}
We assume that our model cloud is supported by pressure alone in equilibrium against the gravity for simplicity.
This assumption means that the magnetic field is assumed to be uniform.
This  may a crude assumption and the magnetic field is likely to be concentrated 
in a dense cloud.  However, it is very difficult to take account of non-uniform magnetic field
\citep[see, e.g.,][]{tomisaka14}.  The observed radial volume density profile is also derived from 
the projected surface density under the assumption that the cloud is symmetric around
the axis.   Thus, it is worth to analyze this very simplified model. 

Since the pressure gradient is given by
\begin{eqnarray}
\frac{dP_0}{dr} & = & \rho _0 g _r , \label{PEQ}
\end{eqnarray}
the pressure and the density should satisfy the relation,
\begin{eqnarray}
\frac{d P _0}{d\rho _0} = \left( \frac{d P _0}{dr} \right) \left( \frac{d\rho _0}{dr} \right) ^{-1} .
\label{EOS}
\end{eqnarray}
When $ p = 2 $, the right-hand side of equation (\ref{EOS}) is evaluated to be
\begin{eqnarray}
\frac{dP _0}{d\rho _0} & =  & \frac{4 \pi G H ^2 \rho _c ^2}{\rho _c - \rho _0} \ln \left( \frac{\rho _c}{\rho _0} \right) , 
\label{EOSp2}
\end{eqnarray}
where equations (\ref{plummer}) and (\ref{gravity}) are substituted into equation (\ref{PEQ}).
Otherwise, it is evaluated to be
\begin{eqnarray}
\frac{dP _0}{d\rho _0} & = & 
\frac{8\pi  G \rho _c H ^2}{p -2} \left( \frac{2 p H ^2}{r ^2} \right) \left( 1 + \frac{r ^2}{2 p H ^2} \right) 
\left[ 1 - \left( 1 + \frac{r ^2}{2 p H ^2} \right) ^{1-p/2} \right] \nonumber \\
& = & \frac{8 \pi G \rho _c H ^2}{p - 2} \left[  1 - \left( \frac{\rho}{\rho _c} \right) ^{2/p} 
\right] ^{-1}  
 \left[ 1 - \left( \frac{\rho}{\rho _c} \right) ^{1-2/p}  \right] . \label{EOSg}
\end{eqnarray}

\begin{figure}[h]
\epsscale{0.5}
\plotone{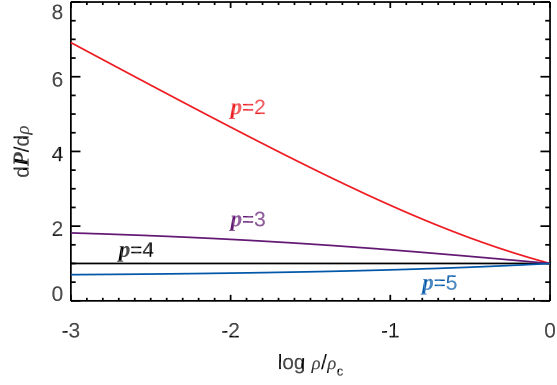}
\caption{The sound speed squared, $ (dP/d\rho) $, is shown as a function of 
$\log (\rho / \rho) $ for $ p $ = 2, 3, 4, and 5 in units of $ 4 \pi G \rho _c H ^2 $. 
\explain{We have added the curve for $ p = 5 $.}
\label{EOS_p}}
\end{figure}

In the following, we assume that Equations (\ref{EOSp2}) and (\ref{EOSg}) hold not only in the equilibrium 
but also for a perturbation.  Thus we use the symbol $ dP/d\rho $ instead of $ dP _0/d\rho _0 $.
Figure~\ref{EOS_p} shows the value as a function of $ \log \left( \rho /\rho _c \right) $
in unit of $ 4 \pi G \rho _c H ^2 $.   
The sound speed ($ \sqrt{dP/d\rho} $) decreases with increasing density for $ p < 4 $, 
while it increases for $ p > 4$.    
In this paper we restrict ourselves to the case of $ p \le 4 $, since the velocity
turbulence is lower in a region of higher density in the interstellar medium.
Thus it is similar to the logatrope proposed by \cite{mclaughlin96}.
They introduced an effective equation of state,
\begin{eqnarray}
P & = & P _c \left[ 1 + \kappa \ln \left( \frac{\rho}{\rho _c} \right) \right] , 
\label{logatrope}
\end{eqnarray}
to mimic interstellar turbulence, where $ P _c $, $ \rho _c $, and $\kappa $ denote
model parameters.   However, the dependence of the sound speed on the density is
weaker than that for logatrope, since Equation (\ref{logatrope}) means
\begin{eqnarray}
\frac{dP}{d\rho} & = & \kappa \frac{P _c}{\rho} .  \label{logatrope2}
\end{eqnarray}

We examine  equation (\ref{EOSg}) again in \S 4.  

We assume that the magnetic field is uniform and runs in the $ x $-direction in the equilibrium.
To specify the initial magnetic field strength in the analysis, we use the plasma beta at the cloud center,
\begin{eqnarray}
\beta & = & \frac{8 \pi \rho _c}{B _0 ^2} \left( \frac{dp}{d\rho} \right) _{\rho = \rho _c} 
= \frac{32 \pi G \rho _c ^2 H ^2}{B _0 ^2} ,
\end{eqnarray}
as in paper I.   
The plasma beta is related to the
mass to flux ratio,
\begin{eqnarray}
f & = & \frac{\displaystyle \int \rho _0 (x ^\prime, y, z) dx ^\prime}{B _0} 
\; = \; \frac{\sqrt{2 \pi p} \Gamma \displaystyle \left( \frac{p-1}{2} \right)}{ \Gamma \displaystyle \left( \frac{p}{2} \right)} 
\left(1 + \frac{y ^2}{2 p H ^2} \right) ^{(-p+1)/2}  \frac{\rho _c H}{B _0} , 
\end{eqnarray}
where $ \Gamma $ denotes the gamma function.  When the mass to flux ratio is critical, i.e.,
$ f _c = \left(2\pi \sqrt{G} \right) ^{-1} $, the plasma beta is
$ \beta =  2/\pi ^2 $, $ 1/3 $, and $ 4/\pi ^2 $ for $ p $ = 2, 3, and 4, respectively.

\subsection{Perturbation Equation}

Following paper I, we consider a small perturbation around the equilibrium in order to 
search for an unstable mode.  
The perturbation is described by the displacement defined by  
\begin{eqnarray}
\mbox{\boldmath$\xi$} & = & e ^{\sigma t} \left[ \xi _{x} (x, y) \cos k z \mbox{\boldmath${e}$} _x +
\xi _y (x, y) \cos k z \mbox{\boldmath$e$} _y + \xi  _z (x, y)  \sin k z \mbox{\boldmath$e$} _z \right] , \label{displacement} 
\end{eqnarray}
where the perturbation is assumed to be sinusoidal in the $ z $-direction with the
wavenumber $ k $ and to grow exponentially with time at the rate, $ \sigma $.  
The change in the density is described as
\begin{eqnarray}
\rho (x, y, z, t) & = & \rho _0 (x, y) + e ^{\sigma t} \delta \varrho (x, y) \cos k z .  \label{delrho1}
\end{eqnarray}
Substituting $ \mbox{\boldmath$v$} = d\mbox{\boldmath$\xi$}/dt $ and Equation (\ref{displacement})
into Equation (\ref{continuity}) we obtain
\begin{eqnarray}
\delta \varrho & = & - \frac{\partial}{\partial x} \left( \rho _0 \xi _x \right) -
\frac{\partial}{\partial y} \left( \rho _0 \xi _y \right) - k \rho _0 \xi _z .  \label{delrho2}
\label{deltarho}
\end{eqnarray}
Similarly, we obtain the perturbation in the magnetic field from the induction
equation,
\begin{equation}
\delta \mbox{\boldmath$B$}  = \mbox{\boldmath$\nabla$} \times 
\left( \mbox{\boldmath$\xi$} \times \mbox{\boldmath$B$} _0 \right) .
\end{equation}
The induction equation  is further expressed as
\begin{eqnarray}
\delta \mbox{\boldmath$B$} (x, y, z) & = & b _x (x, y) \cos k z \mbox{\boldmath$e$}_x
 +  b _y (x, y) \cos k z \mbox{\boldmath$e$} _y + b _z (x, y) \sin k z \mbox{\boldmath$e$} _z  , \\
b _x & = & - B _0 \left(  \frac{\partial}{\partial y} \xi _y  + k \xi _z \right), \label{defbx}\\
b _y & = & B _0 \frac{\partial \xi _y}{\partial x} , \label{defby} \\
b _z & = & B _0 \frac{\partial \xi _z}{\partial x} . \label{defbz}
\end{eqnarray}
We evaluate the change in the current density to be
\begin{equation}
\delta \mbox{\boldmath$J$}  =  \frac{1}{4\pi} \mbox{\boldmath$\nabla$}
\times \delta \mbox{\boldmath$B$} ,
\end{equation}
using Equation (\ref{e-current}).  Each component of the current density is expressed as
\begin{eqnarray}
\delta \mbox{\boldmath$J$} (x, y, z)  & = & j _x (x,y) \sin k z \mbox{\boldmath$e$} _x +
 j _y (x,y) \sin k z \mbox{\boldmath$e$} _y  + j _z (x, y) \cos k z \mbox{\boldmath$e$} _z , \\
 j _x  & = & \frac{1}{4 \pi} \left( \frac{\partial b _z}{\partial y} + k b _y \right) ,  \\
 j _y  & = & - \frac{1}{4 \pi} \left( k  b _x + \frac{\partial b _z}{\partial x} \right) , \\
 j _z  & = & \frac{1}{4 \pi} \left( \frac{\partial b _y}{\partial x} 
 - \frac{\partial b _x}{\partial y} \right) .
\end{eqnarray}
Then the changes in the density and current density are expressed as an explicit function of
$ \mbox{\boldmath$\xi$} $.

The change in the gravitational potential is given as the solution of the Poisson equation
\begin{equation}
\mbox{\boldmath$\nabla$} ^2 \delta \psi = 4 \pi G \delta \rho . \label{poisson}
\end{equation}
Thus, it can be regarded as an implicit function of $ \mbox{\boldmath$\xi$} $.

We derive the equation of motion for the perturbation by taking account of the
force balance,
\begin{equation}
\left( \frac{dP}{d\rho} \right) \mbox{\boldmath$\nabla$} \rho _0 + \rho _0 
\mbox{\boldmath$\nabla$} \psi _0 = 0, 
\end{equation}
with no electric current density, $ \mbox{\boldmath$j$} _0 = 0 $,
in equilibrium.
Then the equation of motion is expressed as
\begin{equation}
\sigma ^2 \rho _0 \mbox{\boldmath$\xi$} = -  \rho _0 
\mbox{\boldmath$\nabla$} \left( \frac{dP}{d\rho} \frac{\delta \rho}{\rho _0} \right)
 -  \rho _0 \nabla \delta \psi 
+ \delta \mbox{\boldmath$J$} \times \mbox{\boldmath$B$} _0 , \label{motion}
\end{equation}
where the last term represents the magnetic force.  The term, $ \mbox{\boldmath$J$} _0
\times \delta \bm{\boldmath$B$} $, does not appear in Equation (\ref{motion}) since 
$ \mbox{\boldmath$J$} _0 = 0 $ in our equilibrium model.
The linear growth rate, $ \sigma $,
is obtained as the eigenvalue of the differential equation (\ref{motion}), since the
right-hand side  is proportional to $ \mbox{\boldmath$\xi$} $.

The derived perturbation equations are the same as those derived in paper I 
except for the sound speed ($ \sqrt{dP/d\rho} $), which is a function of
the density in our analysis but constant in paper I.

\begin{table}[h]
\caption{Variables Describing Perturbations\label{symmetry}}
\begin{center}
\begin{tabular}{llcc}
\hline
variable  & evaluation  & symmetry &  symmetry \\
& point &  $ x $ & $ y $ \\
\hline
$ \xi _x  $ & $ (i-1/2,j) $ & A & S\\
$ \xi _y  $ & $ (i,j-1/2) $  & S & A \\
$ \xi _z  $ & $ (i,j) $ &   S & S \\
$ \delta \varrho$ & $ (i,j) $ &   S & S \\
$ \delta \psi $ & $ (i,j) $ &   S & S \\
$ b_x  $ & $ (i,j) $ &   S & S \\
$ b _y  $ & $ (i-1/2,j-1/2) $ & A & A \\
$ b _z  $ & $ (i-1/2,j) $  & A & S \\
$ j _y  $ & $ (i,j) $ &  S & S \\
$ j _z  $ & $ (i,j-1/2) $ & S & A  \\
\hline
\end{tabular}
\end{center}
\tablecomments{A: anti-symmetric. S: 
symmetric.}
\end{table}

Our equilibrium model is symmetric with respect to the $ x $- and $ y $-axes.   Thus, all 
eigenmodes should be either symmetric or anti-symmetric with respect to these axes.
We restrict ourselves to the eigenmodes symmetric to both $ x $- and 
$ y $-axes, since the unstable mode has the same symmetry in the case of
no magnetic field \citep{nakamura93}.  The choice of this symmetry is justified since
we are interested only in the unstable mode.    Using this symmetry, we can reduce the
region of computation to the first quadrant, $ x \ge 0 $ and $ y \ge 0 $.
The variables describing the perturbation and their symmetries are summarized in Table 
\ref{symmetry}. 

We consider two types of the boundary conditions.   The first one assumes that  
the displacement should vanish in the region very far from the filament center.  
We call this the fixed boundary since the magnetic field lines are fixed on the boundary.  
The second one allows the magnetic field lines to move while remaining straight and 
normal to the boundary.   This restriction is expressed as 
\begin{eqnarray}
\left( \mbox{\boldmath$B$} _0 \cdot \mbox{\boldmath$\nabla$} \right) \mbox{\boldmath$\xi$} 
= 0 .
\end{eqnarray}
Thus, we assume $ \partial \mbox{\boldmath$\xi$} / \partial x $ on the boundary in
the $ x $-direction and $ \mbox{\boldmath$\xi$} = 0 $ in the $ y $-direction.
We refer to this as the free boundary condition.   In both types of boundary conditions, we use the symmetries
given in Table \ref{symmetry} 
to set the boundary conditions for $ x = 0 $ and $ y = 0 $.

\subsection{Numerical Methods}

We solve the eigenvalue problem numerically by a finite difference approach.  The differential
equations are evaluated on the rectangular grid in the $xy$ plane.   We evaluate $ \xi _z $,
$ \delta \varrho $, $ \delta \psi $, $ b _x $,  and $ j _y $ at the points
\begin{eqnarray}
\left( x _i, y _j \right) & = & \left( i \Delta x, j \Delta y \right) ,
\end{eqnarray}
where $ i $ and $ j $ specify the grid points, while $ \Delta x $ and $ \Delta y $ denote 
the grid spacing in the $ x $- and $ y $-directions, respectively (see Table \ref{symmetry}).   
These variables are symmetric with respect to both the $ x $- and $ y $-axes.   Using this
symmetry, we consider the range $ 0 \le i \le n _x $ and $ 0 \le j \le  n _y $, where
$ n _x $ and $ n _y $ specify the number of grid points in each direction.  
When $ i > n _x $ or $ j > n _y $, the displacement
$ \xi _{z,i,j} $ is assumed to vanish  for
the fixed boundary and to have the same values at neighboring points
in the computation domain for the free boundary condition.    
We use the indexes,
$ i $ and $ j $, to specify the position where the variables are evaluated, such
as $ \xi _{z,i,j} = \xi _z (x _i, y _j) $.  

The variables are evaluated at either of
\begin{eqnarray}
\left( x _{i-1/2}, y _j \right) & = & \left[ \left( i - \frac{1}{2} \right)  \Delta x, j \Delta y \right] , \\
\left( x _i, y _{j-1/2} \right) & = & \left[ i \Delta x, \left( j - \frac{1}{2} \right) \Delta y \right] , \\
\left( x _{i-1/2}, y _{j-1/2} \right) & = & \left[\left( i - \frac{1}{2} \right) \Delta x,
 \left( j - \frac{1}{2} \right) \Delta y \right] .
\end{eqnarray}
depending on the symmetry as summarized in Table~\ref{symmetry}.  Here
the symbols, $ i $ and $ j $, are integers to specify the grid points while $ \Delta x $ and $ \Delta y $
denote the grid spacings in the $ x $- and $y$-directions, respectively.
All these variables are evaluated in the region $ 0 \le x \le n _x \Delta x $ and
$ 0 \le y \le n _y \Delta y $.
In other words, we use staggered this kind of grids to achieve second-order accuracy in space.

Using the variables defined on the grids, we rewrite the perturbation equations.
Equation (\ref{deltarho}) is rewritten as
\begin{eqnarray}
\delta \varrho _{i,j} & = &  - \frac{\rho _{0,i+1/2,j} \xi _{x,i+1/2,j} - \rho _{0,i-1/2,j} \xi _{x,i-1/2,j}}{\Delta x} 
 - \frac{\rho _{0,i,j+1/2} \xi _{y,i,j+1/2} - \rho _{0,i,j-1/2} \xi _{y,i,j-1/2}}{\Delta y}  -  k \rho _{0,i,j} \xi _{z,i,j} . \label{continuity3}
\end{eqnarray}
Equation (\ref{poisson}), the Poisson equation, is expressed as
\begin{eqnarray}
\frac{\delta \psi _{i+1,j} + \delta \psi _{i-1,j}}{\Delta x ^2}  
+ \frac{\delta \psi _{i,j+1} + \delta \psi _{i,j-1}}{\Delta y ^2}  
- \left( \frac{2}{\Delta x ^2} + \frac{2}{\Delta y ^2} +  k ^2 \right) \delta \psi _{j,k} 
 = 4 \pi G \delta \varrho _{i,j} . \label{poisson2}
\end{eqnarray}
The solution of Equation (\ref{poisson2}) is expressed as
\begin{eqnarray}
\delta \psi _{i,j} & = & \sum _{i^\prime} \sum _{j^\prime}
G _{i,j,i^\prime, j^\prime} \delta \varrho _{i^\prime,j^\prime} , \label{poisson3}
\end{eqnarray}
where $ G _{i,j,i^\prime,j^\prime} $ denotes the Green's function and
the value is obtained by solving Equation (\ref{poisson2}) numerically.
The boundary condition for the Poisson equation is improved for increasing the accuracy of 
the growth at a small $ k $.  See Appendix \ref{BesselK}  for more details.

The change in the magnetic field  is
evaluated as
\begin{eqnarray}
b _{x,i,j} & = & - B _0 \left( \frac{\xi _{y,i,j+1/2} - \xi _{y,i,j-1/2}}{\Delta y}
+ k \xi _{z,i,j} \right) , 
\end{eqnarray}
\begin{eqnarray}
b _{y,i-1/2,j-1/2} & = & B _0 \left( \frac{\xi _{y,i,j-1/2} - \xi _{y,i-1,j-1/2}}{\Delta x} \right) , \\
b _{z,i-1/2,j} & = & B _0 \left( \frac{\xi _{z,i,j} - \xi _{z,i-1,j}}{\Delta x} \right) ,
\end{eqnarray}
from Equations (\ref{defbx}) through
(\ref{defbz}).
The current density is evaluated as
\begin{eqnarray}
j _{y,i,j} & = & - \frac{1}{4 \pi} \left( k b _{x,i,j} + \frac{b _{z,i+1/2,j} - b _{z,i-1/2,j}}{\Delta x} \right) , 
\end{eqnarray}
\begin{eqnarray}
j _{z,i,j-1/2} & = & \frac{1}{4 \pi} \left( \frac{b _{y,i+1/2,j-1/2} - b _{y,i-1/2,j-1/2}}{\Delta x} 
  -  \frac{b _{x,i,j} - b _{x,i,j-1}}{\Delta y} \right) . \label{jz}
\end{eqnarray}
The $ x $-component of the current density, $ j _x $, is not evaluated, since it does not appear in the
equation of motion.
The fixed boundary conditions are expressed as
\begin{eqnarray}
\xi _{x,n_x +1/2,j} & = & 0 , \\
\xi _{y,n_x +1, j-1/2} & = & 0 , \\
\xi _{z,n_x+1,j} & = & 0 , \\
\xi _{x,i-1/2, n_y+1} & = & 0 , \\
\xi _{y,i,n_y+1/2 } & = & 0 , \\
\xi _{z,i,n_y+1} & = & 0 .
\end{eqnarray}
When the free boundary is applied, the conditions are replaced with
\begin{eqnarray}
\xi _{x,n_x+1/2,j} & = & \xi _{x,n_x-1/2,j} , \\
\xi _{y,n_x+1,j-1/2} & = & \xi _{y,n_x,j-1/2} , \\
\xi _{z,n_x+1,j} & = & \xi _{z,n_x,j} , \\
\xi _{x,i-1/2, n_y+1} & = & \xi _{x,i-1/2, n_y} , \\
\xi _{y,i,n_y+1/2 } & = & \xi _{y,i,n_y-1/2} , \\
\xi _{z,i,n_y+1} & = & \xi _{z,i,n_y} .
\end{eqnarray}

The equation of motion (\ref{motion}) is expressed as
\begin{eqnarray}
\sigma ^2 \rho _{0,i-1/2,j} \xi _{x,i-1/2,j} & = & - \frac{\rho _{0,i-1/2,j}}{\Delta x} 
\left[ \left( \frac{dP}{d\rho} \right) _{i,j} \frac{\delta \varrho _{i,j}}{\rho _{0,i,j}} - 
\left( \frac{dP}{d\rho} \right) _{i-1,j} \frac{\delta \varrho _{i-1,j}}{\rho _{0,i-1,j}} \right]
 - \frac{\rho _{0,i-1/2,j}}{\Delta x} \left( \delta \psi _{i,j} - \delta \psi _{i-1,j} \right) . \label{motion1} \\
 \sigma ^2 \rho _{0,i,j-1/2} \xi _{y,i,j-1/2} & = & 
- \frac{\rho _{0,i,j-1/2}}{\Delta y} 
\left[ \left( \frac{dP}{d\rho} \right) _{i,j} \frac{\delta \varrho _{i,j}}{\rho _{0,i,j}}  
-  \left( \frac{dP}{d\rho} \right) _{i,j-1} \frac{\delta \varrho _{i,j-1}}{\rho _{0,i,j-1}} \right]
- \frac{\rho _{0,i,j-1/2}}{\Delta y} \left( \delta \psi _{i,j} - \delta \psi _{i,j-1} \right)  \nonumber \\
& \; &  + B _0 j _{z,i,j-1/2}  . \label{motion2} \\
\sigma ^2 \rho _{0,i,j} \xi _{z,i,j} & = & - k \left( \frac{dP}{d\rho} \right) _{i,j} \delta \varrho _{i,j} 
- k \rho _{0,i,j} \delta \psi _{i,j}   - B _0 j _{y,i,j} . \label{motion3}
\end{eqnarray}

Equations (\ref{motion1}) through (\ref{motion3}) are summarized in the form,
\begin{eqnarray}
\sigma ^2 \mbox{\boldmath$B$} \mbox{\boldmath$\zeta$} = \left( \mbox{\boldmath$A$} + B _0 ^2
\mbox{\boldmath$C$} \right) \mbox{\boldmath$\zeta$} , \label{algebraic}
\end{eqnarray}
by using Equations (\ref{continuity3}), and (\ref{poisson3})  through (\ref{jz}).  Here,
$ \mbox{\boldmath$\zeta$} $ denotes an array of components, 
$ \xi _{x,i-1/2,j} $, $ \xi _{y,i,j-1/2} $, and $ \xi _{z,i,j} $ for all the combinations of $ i $ and $ j $.
The matrix elements of $ \mbox{\boldmath$A$} $, $ \mbox{\boldmath$B$} $, and
$ \mbox{\boldmath$C$} $ are evaluated numerically as a function of $ k $.   
See Appendix B of Paper I for further details.   
Then the growth rate is given as the solution of 
\begin{eqnarray}
\det \left(  \sigma ^2 \mbox{\boldmath$B$} - \mbox{\boldmath$A$} - B _0 ^2 
\mbox{\boldmath$C$} \right) = 0 . \label{generalE}
\end{eqnarray}
We rewrite Equation (\ref{generalE}) into
\begin{eqnarray}
\det \left[ \sigma ^2 - \mbox{\boldmath$B$} ^{-1/2} 
\left(  \mbox{\boldmath$A$} - B _0 ^2 
\mbox{\boldmath$C$} \right) \mbox{\boldmath$B$} ^{-1/2} \right] = 0 ,
\label{generalE2}
\end{eqnarray}
where matrix, $\mbox{\boldmath$B$} ^{-1/2} $, is obtained easily since only the
diagonal elements have non-zero values in  matrix, $\mbox{\boldmath$B$} $. 
Equation (\ref{generalE2}) is an eigenvalue problem while equation (\ref{generalE})
is a generalized eigenvalue problem.   Various library programs are available for
solving the former. 
We use  subroutine DGEEVX of LAPACK 
\citep[see,][for the software]{anderson99} to solve Equation (\ref{generalE2}).
The subroutine returns all the eigenvalues $ \sigma ^2 $.     

The matrixes $ \mbox{\boldmath$A$} $, $ \mbox{\boldmath$B$} $, and
$ \mbox{\boldmath$C$} $ have 
dimension $ \left( 3 n _x n _y + 2 n _x + 2 n _y + 1 \right) $. Thus, we obtain
$ 3 n _x n _y + 2 n _x + 2 n _y + 1 $ eigenmodes.  However, we select only
one unstable mode ($ \sigma ^2 > 10 ^{-5} $) for a given $ k $ and $ B _0 $.
The remaining eigenmodes denote oscillation of the filamentary cloud.
In the following, we restrict ourselves to the unstable mode.  

When $ B _0 = 0 $, our equilibrium model is symmetric around the axis and we can
simplify the stability analysis using the cylindrical coordinates.  The numerical
methods are summarized in Appendix \ref{1Dmode}.

When $ k = 0 $, we need not solve Equation (\ref{motion3}), since 
the $ z $-component of the displacement, $ \xi _z $, vanishes.  Accordingly,
we can omit the corresponding part of the matrix given Equation (\ref{generalE2}).
The dimension of the matrix to be solved reduces to $ 2 n _x n _y + n _x + n _y $.
The boundary condition for the Poisson equation is given at  the end of Appendix \ref{BesselK}.

\section{RESULTS}

\subsection{Case of $ B _0 = 0$}

\begin{figure}[h]
\epsscale{0.5}
\plotone{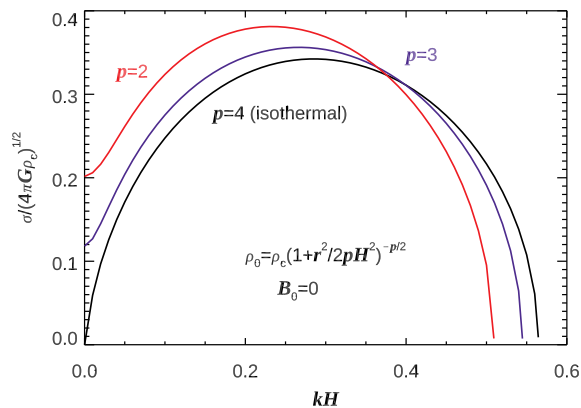}
\caption{Growth rate is shown as a function of the wavenumber
for $ p $ = 2, 3, and 4. \label{NoBgrowth}}
\end{figure}

Before examining the effects of magnetic field, we analyze the stability of our Plummer-like
model for  the case of $ B _0 = 0 $.    When the magnetic field vanishes, our equilibrium
model is symmetric around the $ z $-axis.  Hence the stability analysis is reduced to 1D
problem.   We obtained the growth rate, $ \sigma $, as a function of the
wavenumber, $ k $, for $ p = 2$, 3, and 4 according to the method given in
Appendix  \ref{1Dmode}.   The growth rate is obtained by solving the discretized 
perturbation equation with the spatial resolution, $ \Delta r = 0.1 H $ and
the boundary condition at $ r _{\rm out} = 60 H $.  Thus the obtained growth
rate is highly accurate.   Figure~\ref{NoBgrowth} denotes the growth rate
in unit of $ \sqrt{4\pi G \rho _0} $ with the wavenumber resolution, $ \Delta k  = 0.01 H^{-1}$.

Similar to the isothermal model, the Plummer models of $ p = 2 $ and 3 are unstable
against fragmentation when the wavenumber is smaller than the critical one.
The critical wavenumber, $ k _{\rm cr} $, is smaller for a lower index, $ p $, when
measured in unit of $ H ^{-1} $.  It is $ k _{\rm cr} = 0.509~H^{-1} $, $ 0.545~H^{-1} $
and $ 0.565~H^{-1} $ for $ p $ = 2, 3, and 4, respectively.
Also the wavenumber for which the growth rate
takes its maximum value is also smaller for a lower index.  However, it is premature
to conclude that a filamentary cloud tends to fragment with a longer interval since
the difference is small.  We need to define the diameter of the filamentary cloud 
more carefully for obtaining 
before quantitative conclusion.  It should be reminded that the radial density profile is
broader for a lower $ p $ when $ \rho _c $ and $ H $ are fixed.
As shown in Figure~\ref{r-density}, the diameter of $ p = 2 $ model  is 
slightly larger than that of $ p = 4 $ if the filament diameter is defined
as the full width at the half maximum.

The maximum growth rate,  $ \sigma _{\rm max} $, is higher for a lower, $ p $, 
when measured in the unit of $ \sqrt{4 \pi G \rho _c} $.  
Again the dependence of $ \sigma _{\rm max} $ on  $ p $ is weak.  This is likely
due to the fact that the cloud is more massive than the isothermal cloud when $ \rho _c $ and $ H $ are
fixed, c.f., Equation (\ref{linemass}). 
 
Lowering the index induces radial collapse of the filament.
When $ p < 4 $, the model is unstable at $ k H = 0 $, i.e., against radial collapse.
This is because the effective sound speed decreases as the density increases.
Remember that the critical line mass is  $ \lambda _{\rm cr} = 2 c _s ^2 / G $ for
 an isothermal  filamentary to be sustained by gas pressure against collapse.
 The line mass of {our} equilibrium model is evaluated to be 
 \begin{eqnarray}
 \lambda _{\rm eq} & = & \frac{4 \pi G \rho _c H ^2}{p - 2} , \label{linemass2}
 \end{eqnarray}
 for $ 2 < p < 4 $ from equation (\ref{linemass}).   Equation (\ref{linemass2}) means
 that the line mass is proportional to the square of the effective sound speed. Thus
 the radial collapse is stabilzed only when the effective sound speed increases as the
 density increases.  Otherwise, the self-gravity overwhelms the gas pressure and
 the radial collapse sets in.  
 
 \subsection{Case of $ k H = 0 $}

\begin{figure}[h]
\epsscale{0.5}
\plotone{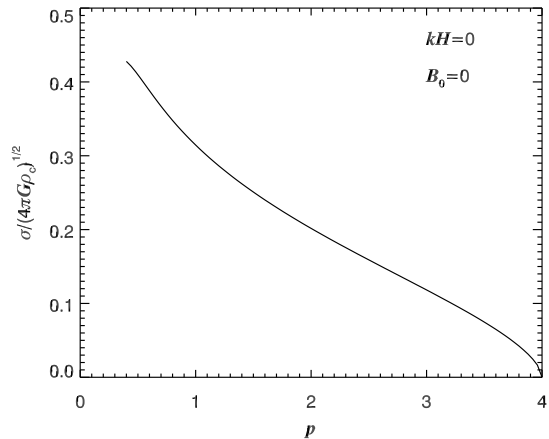}
\caption{Growth rate is shown as a function of the index, $ p $, for
for radial collapse, $ k = 0 $ of the unmagnetized model ($ B _0 = 0 $). \label{1Dgrowth}}
\end{figure}

 Figure \ref{1Dgrowth} denotes the growth rate of the radial collapse  ($ k H = 0 $) mode, 
 $ \sigma/\sqrt{4\pi G \rho _c} $, as a function of the index, $ p $, for $ B _0 = 0 $.  
 It is obtained numerically with the method shown in Appendix \ref{1Dmode} with
the outer boundary at $ r =  200 H $.   
The growth rate is lower for a higher $ p $ and vanishes at $ p = 4 $ (isothermal).    
It should be also noted we find only one unstable mode for a given $ k H $.

This dependence of the instability on the equation of sate has been
well known as summarized in the review by \cite{larson03}.  Recently,
\cite{toci15a} have reported a similar result on the radial collapse of a filamentary cloud.
They assumed the polytropic equation of state, $ P = K \rho ^{\gamma _{\rm p}} $, 
where $ K $ and $ \gamma _{\rm p} $ are a constant and the polytropic exponent, respectively.
When $ \gamma _{\rm p} < 1 $, their model cloud is also unstable against radial collapse.
When $ \gamma _{\rm p} > 1 $, the model cloud is stable against radial collapse and
the density vanishes at a finite radius. 

\begin{figure}[h]
\epsscale{0.5}
\plotone{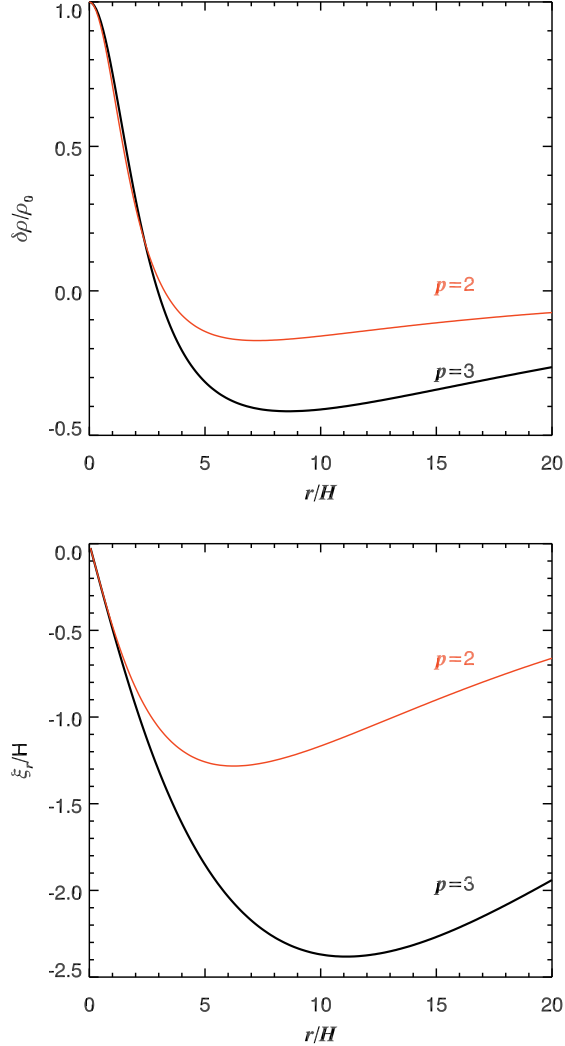}
\caption{The radial collapse mode is shown as a function of $ r $.   The upper panel
shows the relative density perturbation while the lower one does the radial displacement.
The red curves denote the eigenmode of $ p = 2 $ while the black ones do those of $ p = 3$.
The amplitude of the eigenmode is normalized so that $ \delta \varrho / \rho _0 = 1 $ at $ r = 0$.
\label{Eigen_p23}}
\end{figure}

The dependence of the growth rate on the index, $ p $, is moderate while it is
larger for a lower $ kH $.   Thus we examine the eigenmode of the radial
collapse for $ p = 2 $ and 3.   The upper panel of Figure \ref{Eigen_p23} shows the relative
density perturbation, $ \delta \rho / \rho _0 $,  as a function of $ r $ for the radial collapse mode,  $ k H = 0 $
and $ B _0 = 0 $.   The lower panel of Figure \ref{Eigen_p23} shows the radial displacement, $ \xi _r /H $.
The red curve denotes the eigenmode of $ p =  2 $ while
the black curve does that of $ p = 3 $.   The eigenmodes are normalized
so that the relative density perturbation be unity, $ \delta \varrho / \rho _0 = 1 $.
See Equations (\ref{delrho1}) and (\ref{delrho2}) for the definition of $\delta \varrho $
and its relation to the displacement.

When the index is smaller than $ p < 4 $, the effective equation of state is \lq \lq soft'' 
in a sense that  the effective sound speed decreases as the density increases.  Thus the
filamentary cloud is subject to the radial collapse.  When $ p = 2 $ and 3, the radial displacement
has a maximum at  $ r \simeq 5 H$ and $ 11 H $, respectively.  When $ p $ is small, an inner part 
region around the axis collapses radially.  When $ p $ is close to 4, the radial collapse
is realized only when the whole cloud collapses in the radial direction.

\begin{figure}[h]
\epsscale{0.5}
\plotone{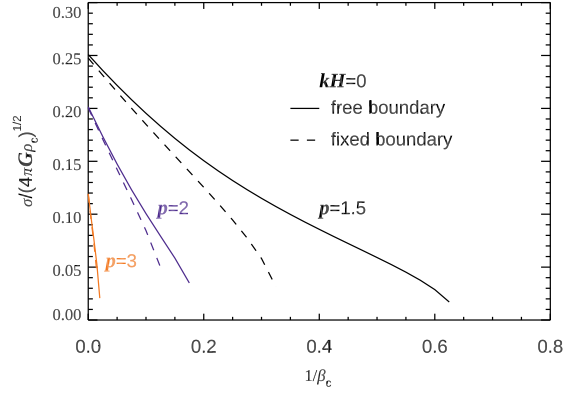} 
\caption{The growth rate of $kH =0 $ mode is shown as a function of $ 1/ \beta _{\rm c} $
for $ p = 1.5$, 2, and 3.  The solid curves denote those for the free boundary
while the dashed ones those for the fixed boundary at $ | x | = 32 H $. \label{kH0B}}
\end{figure}

The radial collapse ($ k H = 0 $) is suppressed by a relatively weak magnetic field.
Figure \ref{kH0B} shows the growth rate as a function of the inverse of the plasma
beta, i.e., the magnetic pressure normalized by the gas pressure at the cloud center.
The solid curves denote the growth rates for the free boundary, while the  dashed
ones those for the fixed boundary at $ x = 32 H$.  The index is set to be $ p = $ 1.5, 2 and 3. 
Relatively weak magnetic field of $ \beta _{\rm c} = 5 $ suppresses the $ kH = 0 $ mode
for $ p = 2 $.  When $ p = 3 $, the radial collapse mode is completely suppressed
by very weak magnetic field of $ \beta _{\rm c}  = 40 $.

The magnetic field stiffens the equation of state since the magnetic pressure increases
more steeply than the gas pressure when compressed.   The radial collapse is
thought to be suppressed when the equation of state is 'isothermal' in effect, 
i.e., when the effective sound speed changes little by the increase in the density.

The growth is lower for given $ p $ and $ \beta $ when the fixed boundary condition
is applied.  When the magnetic field is fixed in the region far from the cloud, the 
magnetic tension works against the collapse in addition to the magnetic pressure.
The difference is larger for a smaller $ p $.  The difference is negligibly small for
$ p $  = 3.   When $ p = 2 $, the difference is appreciable for $ \beta < 10 $.  
The difference depends a little on the computation domain.  The growth rate depends little 
on the size of the computation domain ($ n _x $), when the free boundary is applied.
However, the growth rate depends a little on $ n _x $ when the fixed boundary is applied.

\subsection{Case of $ B _0 \ne 0 $ and $ k H \ne 0 $}

First, we examine the case of $ p = 2 $, since the density profile is close to the observed one.

\begin{figure}[h]
\epsscale{0.5}
\plotone{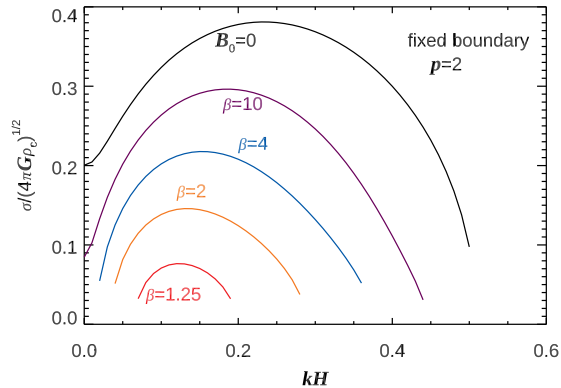}
\caption{Growth rate, $\sigma $, is shown in unit of $\sqrt{4\pi G \rho _{\rm c}} $ 
as a function of the wavenumber, $ kH $,
for $ p  = 2 $ and  the fixed boundary.  \label{DorisFix}}
\end{figure}

Figure \ref{DorisFix} denotes the growth rate, $ \sigma /\sqrt{4 \pi G \rho _c} $, as
a function of the wave number, $ k H $, for $ p = 2 $ and the fixed boundary condition.
Each curve denotes the growth rate for a given $ B _0 $.  The label denotes the
plasma beta on the cloud axis.  Figure \ref{DorisFix} is obtained with the spatial
resolution, $ \Delta x = \Delta y = 0.4 H $, the computation box size, $ n _x = n _y = 80 $,
and the wavelength resolution, $ \Delta k = 0.01 H ^{-1} $.

The growth rate is lower for a stronger magnetic field.   
When the initial magnetic field is uniform as assumed in our model,
any perturbation increases magnetic energy.   Hence, any perturbation
induces restoring force proportional to the square of the magnetic field.
The radial collapse is suppressed for $ \beta < 4 $ since the magnetic field
reduces the compressibility of the gas.
The unstable mode is completely stabilized when $ \beta < 0.7 $, while
the entire cloud is subcritical for $ \beta < 0.203 $ (see \S 2.1).
This result is analogous to that in Paper I.  The isothermal cloud is completely
stabilized for $ \beta < 1.67 $ while the entire cloud is subcritical for $ \beta < 0.405 $.

The growth rate depends a little on the size of the computation domain, 
$ n _x \Delta x $.  The dependence on the domain size is appreciable when
the growth rate is small.
When $ p = 2 $, $ \beta = 2.0 $, $ k H = 0.2 $ and $\Delta x = 0.4~H$, 
the growth rate is 0.124 and 0.139~$\sqrt{4 \pi G \rho _c}$ for $ n _x = 80 $ and 120, respectively.
The initial density is very low $ \rho _0 / \rho _c  = 3.89 \times 10 ^{-3} $ and $1.73 \times 10^{-3} $
at the numerical boundary at $ 32 H $ and $ 48 $, respectively.  However, the location
of the boundary affects the instability through the magnetic tension.  When the magnetic
field is fixed at a relatively short distance, the magnetic tension is strong enough to
stabilize the cloud against fragmentation. 

\begin{figure}[h]
\epsscale{0.5}
\plotone{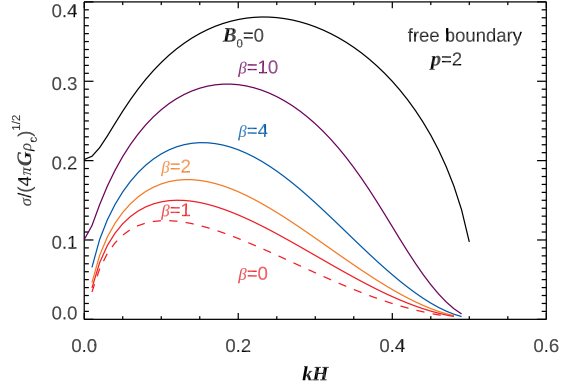}
\caption{The same as Fig. \ref{DorisFix} but for the free boundary. \label{DorisFree}}
\end{figure}

Figure \ref{DorisFree} is the same as Figure \ref{DorisFix}  but for the
free boundary.  The growth rate is obtained with the same resolution, $ \Delta x = \Delta y = 0.4 H $,
$ n _x = n _y = 80 $ and $ \Delta k = 0.01 H ^{-1} $.
When $ \beta  < 1 $, the growth rate is well approximated by an empirical formula,
\begin{eqnarray}
\sigma ^2 (k, \beta) & = & \sigma ^2 (k, 0)  + \frac{d\sigma ^2}{d\beta} 
 \beta + {\cal O} (\beta ^2) ,
\end{eqnarray} 
where $ \sigma (k, 0) $ and $ d\sigma ^2/d\beta $ are positive constants for a given $ k $.
The growth rate for $ \beta = 0 $ shown in Figure \ref{DorisFree} is obtained
by the linear extrapolation of the growth rates at $ \beta = 0.1 $ and 0.5.

The growth rate depends little on the size of the computation domain.
When $ p = 2 $, $\beta = 2 $, $ kH = 0.2 $ and $ \Delta x = 0.4 H $, the growth
rate is 0.160 and 0.161~$\sqrt{4 \pi G \rho _c}$ for $ n _x = 80 $ and 120, respectively.

The growth rate is lower for a stronger magnetic field also when the free boundary is applied.
The radial collapse is suppressed by a mildly strong magnetic field ($ \beta < 4 $).
However, the model cloud is unstable against fragmentation even when the entire cloud is
subcritical.   This instability is due to rearrangement of magnetic flux tubes as shown in
paper I.   Although the critical wavenumber changes little, the wavenumber of the most
unstable mode decreases down to $ (kH) _{\rm max}  \simeq 0.11 $ in the limit of $ \beta = 0 $.

Comparison of Figures \ref{DorisFix} and \ref{DorisFree} tells us that the growth
rate depends significantly on the boundary condition only when $ \beta \la 2 $.  
When the gas pressure dominates over the magnetic pressure on the cloud axis 
($ \beta \ga 2 $), the growth rate is nearly the same for both the free and fixed
boundaries except for $ k H \la 0.4$, i.e., when the wavenumber is close
to $ k _{\rm cr} $ and the instability is weak even for $ B _0 = 0$.   
This means that a relatively weak magnetic field does not play an
important role in the region far from the axis, although the magnetic pressure
dominates over the gas pressure thereof.  Remember that the magnetic
pressure is comparable to the gas pressure at $ r = 12 H $ even when the
$ \beta = 10 $.  The gas pressure decreases with the decrease in the 
density while the magnetic pressure remains constant in our model.  
The magnetic pressure dominates over the gas pressure near the outer boundary
($ x = 32 H $ and $ y  = 32 H $) of our numerical computation even for $ \beta = 100 $. 
The magnetic field can suppress the fragmentation only when the magnetic field is
strong near the cloud axis and fixed in the region very far from the cloud.

The Plummer index, $ p $, can vary from cloud to cloud.   Thus we examine two
cases, $ p = 1.5 $ and $ p = 3 $.   The former is close to the observed minimum.

\begin{figure}[h]
\epsscale{0.5}
\plotone{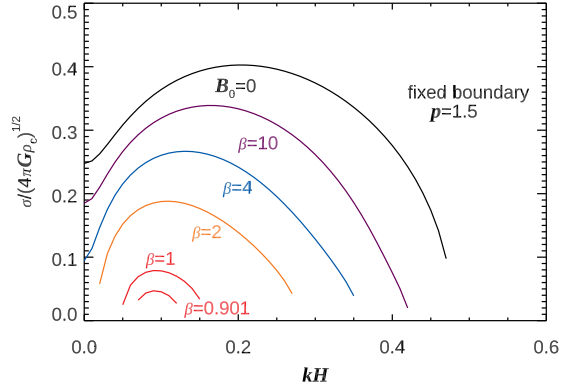}
\caption{The same as Figure \ref{DorisFix} but for $ p  = 1.5 $ and 
the fixed boundary. \label{p15Fix}}
\end{figure}

\begin{figure}[h]
\epsscale{0.5}
\plotone{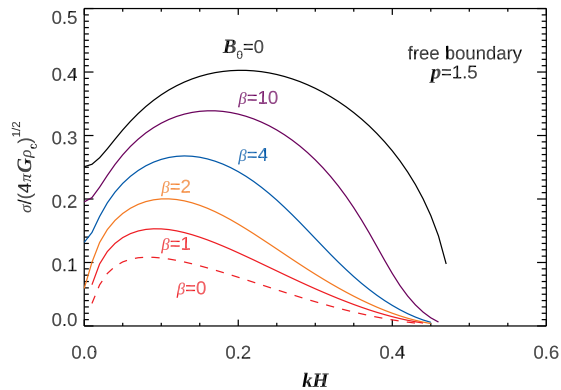}
\caption{The same as Fig. \ref{DorisFree} but for $ p = 1.5 $ and the 
free boundary.  \label{p15Free}}
\end{figure}

Figures \ref{p15Fix} and \ref{p15Free} show the growth rate, $ \sigma /\sqrt{4\pi G \rho _c} $,
as a function of the wavenumber, $ kH $, for the model of $ p = 1.5 $.   Figure \ref{p15Fix} denotes
the growth rate for the fixed boundary while Figure \ref{p15Free} does that for the free boundary.
The results are qualitatively similar to those for $ p = 2 $, while the growth rate is a little larger 
for given $ k H $ and $ \beta $.  As a result, the model cloud is unstable against fragmentation
for $ \beta = 1 $ even when the fixed boundary is applied.  The growth rate depends significantly
on the boundary condition when $ \beta \la 1 $.  

The increase in the growth rate might be due to the normalization.
The growth rate is normalized by the initial central density ($ \rho _c $).  The wavenumber is
normalized not only by  the density but by the sound speed 
($ \sqrt{dP/d\rho} $) on the cloud axis, since the unit length can be expressed as
\begin{eqnarray}
H & = & \frac{1}{\sqrt{8 \pi G \rho _c}} \left[ \frac{dP _0}{d\rho _0} \left(\rho _c \right) 
\right] ^{1/2} .  \label{unitH}
\end{eqnarray}
Equation (\ref{unitH}) is derived from Equations (\ref{EOSp2}) and (\ref{EOSg})
by taking the limit of $ r \rightarrow 0 $.  In other words, the growth rate is
normalized by the free-fall timescale at the cloud center, while the wavenumber is
normalized by the Jeans length, $ \lambda _{\rm J} = 2 \pi H$.
When the central density and sound speed are
fixed, the mass per unit inside radius, $ r $, is larger for a smaller $ p $.  The increase
in the growth rate is likely to be ascribed to the increase in the mass per unit length 
since this instability is due to the self-gravity.


\begin{figure}[h]
\epsscale{0.5}
\plotone{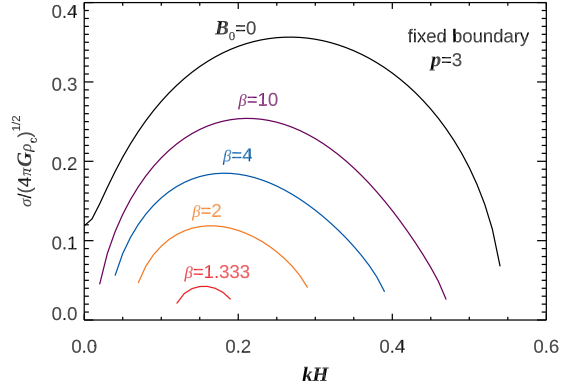}
\caption{Growth rate is shown as a function of the wavenumber
for $ p $ = 3.  The fixed boundary condition is applied.  \label{p3Fix}}
\end{figure}

\begin{figure}[h]
\epsscale{0.5}
\plotone{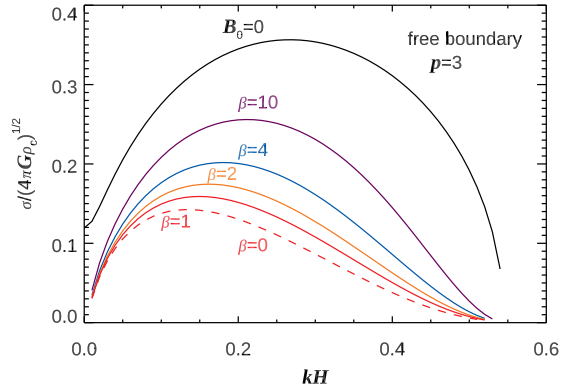}
\caption{The same as Fig. \ref{p3Fix} but for the free boundary.  \label{p3Free}}
\end{figure}

Figures \ref{p3Fix} and \ref{p3Free} are the same as Figures \ref{DorisFix} and
\ref{DorisFree} but for $ p = 3 $, respectively.   The result depends only quantitatively
on $ p $ as expected.  The growth rate is intermediate between those for $ p = 2 $  and 4.
When the fixed boundary is applied, the cloud is stabilized by a moderately strong magnetic field
($ \beta \la 1.3 $).    The radial collapse ($ k H \ll 0.05 $) mode is suppressed by a relatively
weak ($ \beta \approx 10 $) magnetic field.  


The eigenfunction depends only weakly on the index, $ p $.  
We do not find any
qualitative change except for the radial collapse in case of weak magnetic field.

\section{DISCUSSIONS}

First we compare our equilibrium model clouds with earlier theoretical models based on
effective equation of state.

\cite{ostriker64} obtained cylindrical equilibrium models by assuming polytropic equation of state,
\begin{eqnarray}
P & = &  K _N \rho ^{1+1/N} , \label{polytrope}
\end{eqnarray}
where $ K _N $ and $ N $ denotes the polytropic constant and index, respectively.   His equilibrium
model is symmetric around the axis and extended infinitely along the axis.
Since he studied the case  of $ N \ge 1$, the effective sound speed, $ \left( dP/d\rho \right) ^{1/2} $,
increases as the density increases.  In his model, the model cloud is truncated at a certain radius
when $ N $ is finite.   Only when $ N $ is infinite (isothermal), the cloud is extended to an infinite
radius.

When $ N $ is negative, the polytrope gives a similarity solution denoting radial collapse 
\citep{mclaughlin97,kawachi98}.   The density decreases in proportion to
\begin{eqnarray}
\rho & \propto & r ^{-2/(1-1/N)} ,  \label{kawachi}
\end{eqnarray}
in the region far from the cloud axis.  This radial profile is quite similar to the Plummer-like
model with $ p \simeq 2 $, when $ N $ is negatively large.

\cite{mclaughlin97} introduced a model with a slightly different equation of state, Equation (\ref{logatrope}), 
which is named \lq logatrope\rq.  The sound speed is inversely proportional to the density as shown
in Equation (\ref{logatrope2}).  The polytrope gives a singular equilibrium having the density profile, 
$ \rho \propto r ^{-1} $.   This corresponds to the polytrope model of $ N = -1 $.

Note that the polytropic equation of state has a steeper dependence on the density compared
with our model equation of state.  The sound speed is proportional to a power of the density,
$ dP/d\rho \propto \rho ^{1/N} $. 
Our model equation of state shows further weaker dependence
of the sound speed on the density as shown in Figure \ref{EOS_p}.    Note that the effective sound speed is only 2.16 and  1.28 times
 at $ \rho = \rho _c /100 $ than that at $ \rho = \rho _c $ for $ p = 2 $ and 3, respectively.
 The small change produces notable change in the radial density profile.  We introduced this
 effective sound speed by taking account of the turbulence, but physical change in the temperature
 may contribute to density profile.   It should be noted that exactly and almost isothermal models
 give different radial density profiles for the equilibrium.
 
 When the effective sound speed decreases with increase in the density, the cloud is unstable
 against radial collapse.   Our analysis indicates that the radial collapse can be suppressed by
 magnetic field perpendicular to the cloud, if they are appreciable.  The conclusion is likely to
 be valid in general.
 
 Our analysis indicates that perpendicular magnetic field affects the dynamics of a filamentary
 cloud when the plasma beta is close to unity.  
 The plasma beta can be evaluated from the
 effective sound speed ($ c _{\rm s,eff} $), the filament width ($w$), and the magnetic field, $ B _0 $.
 if the filamentary cloud is supported by pressure, the filament mass per unit length is
 evaluated to be
 \begin{eqnarray}
 \lambda & \approx & \frac{2 c _{\rm s,eff} ^2}{G} .  \label{line-mass}
 \end{eqnarray}
Then the gas pressure at the cloud center is evaluated to be
\begin{eqnarray}
P _{\rm c} & \approx & \frac{4 c _{\rm s,eff} ^2 \lambda}{\pi w^2} .
\end{eqnarray}
Thus the plasma beta is evaluated to be
\begin{eqnarray}
\beta & \approx & \left( \frac{c _{\rm s,eff} }{0.3~\mbox{km~s}^{-1}} \right) ^4
\left( \frac{w}{0.1~\mbox{pc}} \right) ^{-2} 
\left( \frac{B _0}{90~\mu\mbox{G}} \right) ^{-2} .  \label{beta-es}
\end{eqnarray}
Equation (\ref{beta-es}) implies magnetic field of $ \sim 100~\mu$G plays an
important role in the dynamics of a typical molecular cloud, since the numbers
quoted are typical.  The plasma beta can be evaluated also from $ w $ and $ \lambda $,
if Equation (\ref{line-mass}) is valid.

It is interesting to apply the above estimate to the Musca filamentary cloud.
The central 1.6 pc of the filamentary cloud shows no sign of fragmentation, although
the rest of the cloud shows fragmentation consistent with stability 
analysis \citep{kainulainen16}.  If the stabilization is due to perpendicular magnetic field,
the field strength should be several tens $\mu$G.  According to \cite{kainulainen16}, 
the best fit Plummer model gives $ p $ = 2.6 and 1.8 in the west and east sides of the filamentary cloud,
respectively.  They also evaluate the effective sound speed to be 22 \% higher than the 
isothermal one.  The right hand side of Equation (\ref{beta-es}) is roughly unity 
when $ B _0 \approx 100~\mu\mbox{G}$. 

Our stability analysis has demonstrated that the instability depends on the boundary
condition, i.e., fixed or free, when the plasma beta is close to unity.   This means that 
a filamentary cloud is not isolated and the stability depends on the environment.    
This suggests an interesting possibility.
If two filamentary clouds are threaded by the same perpendicular magnetic field lines, their
fragmentation can be linked through magnetic force.  Fragmentation of a cloud might
be suppressed since the magnetic field lines are fixed by the other cloud.
However, we cannot exclude the possibility that two clouds fragment coherently.
This problem is beyond the scope of this paper and an open question.

 Except for the radial collapse, the growth rate of the perturbation depends only quantitatively
 on the index, $ p $.   Any strong magnetic field cannot stabilize the cloud if the magnetic fields are 
 free to move in the region far from the cloud.   However, this does not mean that a filamentary
 cloud fragment to form cores in a short timescale.   The growth rate is nearly by a factor of ten
 smaller than the dynamical one, $\sqrt{4\pi G\rho _c} $, for $ p = 2 $ as shown 
 in Figure \ref{DorisFree}.   Remember that the growth rate is normalized by the free-fall
 timescale at the cloud center,
 \begin{eqnarray}
 \tau _{\rm ff} \; = \; \frac{1}{\sqrt{4 \pi G \rho _c}} 
 & = & 0.18~\left( \frac{n _{{\rm H} _2}}{10 ^4~\mbox{cm}^{-3}} \right) ^{-1/2}
 ~\mbox{Myr} .
 \end{eqnarray}
Since the growth rate is small, we need to take account of 
non-ideal effects, i.e., the ambipolar diffusion of magnetic field, which is ignored in our analysis
for simplicity.  It is well known that the ambipolar diffusion increases mass to flux ratio locally and
makes a molecular cloud eventually supercritical.   This competing process should be taken
account seriously, if we discuss the instability due to the rearrangement of magnetic flux tubes.

We also note that the growth rate of the rearrangement instability is smaller for a smaller, $ p $,
although the maximum growth rate of $ B _0 = 0 $ is larger for a smaller $ p $.
This result is consistent with the discussion given in paper I.   Rearrangement of magnetic
fields travel along the field line as the Alfv\'en wave, which is proportional to $ B _0/\sqrt{\rho _0} $.
Thus the rearrangement takes more times when either magnetic field weakens or the density
decreases more slowly in the region far from the cloud.   In our analysis, only the latter effect
is taken into account, while the magnetic fields should be weaker outside the cloud than inside.
Thus the rearrangement instability grows more slowly if non-uniformity of magnetic field is
taken account.

\section{Summary}

We have examined the stability of filamentary cloud permeated by uniform perpendicular magnetic
fields with a focus on the dependence on the initial density.  
Our main findings are summarized as follows.
\begin{enumerate}
\item The observed Plummer-like density profile can be realized if the effective sound speed 
increases with decrease in the density.   The radial density slope is $ d\ln \rho/d \ln r = - 2 $,
if the effective sound speed is only by a factor of 5 larger in the region
where the density is by a factor of 100 lower than at the cloud center.
This dependence of the effective sound speed on the density is much lower than that
of a logatrope. 
\item When the radial density slope is shallower than $ d\ln \rho/d\ln r >  - 4$,
the cloud is unstable against radial collapse.  The growth rate is larger when the
radial density slope is shallower. The radial collapse can be suppressed
by mildly strong magnetic fields.
\item Stability of a filamentary cloud depends strongly on the boundary condition far from
the filament, i.e., on the environment.  If the magnetic field is fixed at a large distance 
from the cloud and the plasma beta is close to unity at the center of the cloud, 
fragmentation is suppressed.   If the magnetic field line can move freely, it cannot
suppress the fragmentation even when the cloud is magnetically subcritical.
The latter instability is induced by rearrangement of magnetic flux tubes as
shown in Paper I.  
\end{enumerate}

\acknowledgements

We thank anonymous referees for constructive comments on the original version of this manuscript.
This work was supported by JSPS KAKENHI Grant Number JP15K05032 and JP 18K03702.   
Most of the numerical computations have been performed on SR24000 at Institute of Management and
Information Technologies, Chiba University.

\software{LAPACK, Linear Package Algebra \citep{anderson99}}

\appendix

\section{Boundary Condition for the Poisson Equation} \label{BesselK}

We improved the boundary condition for the Poisson equation.   
Equation (\ref{poisson3}) relates the change in the gravitational potential  to
that in the density, where the Green's function, $ G _{i,j,i^\prime,j^\prime} $,
takes account of the boundary conditions on $ x = 0 $ and $ y = 0 $.
In this work we removed the boundary by employing the method of the mirror image,
i.e., by taking account of the change in the density in the regions of 
$ x < 0 $ and/or $ y < 0 $.    Then the boundary conditions for the Poisson equations are
expressed as
\begin{eqnarray}
\lim _{\sqrt{x ^2 + y ^2}  \rightarrow \infty} \left| \mbox{\boldmath$\phi$} \right| = 0 .
\end{eqnarray}
The corresponding Green's function is expressed as
\begin{eqnarray}
G ^\prime _{i,j,i^\prime,j^\prime} & = & 2 G \Delta x \Delta y K _0 (k r) ,  \label{modK} \\
r & = & \sqrt{ \left( i - i ^\prime \right) ^2 \Delta x ^2 + \left(j - j ^\prime \right) ^2 \Delta y ^2 } ,
\end{eqnarray}
where $ K _0 $ denotes  the 0-th modified Bessel function of the second kind and
the argument is the distance from the source multiplied by the wavenumber, $ k $.
We solved the discretized Poisson equation,
\begin{eqnarray}
\frac{G ^\prime  _{i+1,j,0,0} - 2 G ^\prime  _{i,j,0,0} + G ^\prime _{i-1,j,0,0}}{\Delta x ^2} + 
\frac{G ^\prime  _{i,j+1,0,0} - 2 G ^\prime  _{i,j,0,0} + G ^\prime _{i,j-1,0,0}}{\Delta y ^2} 
- k ^2 G ^\prime _{i,j,0,0} 
& = & 
\begin{cases}
1 & ( i = j = 0) \\
0 & \mbox{(otherwise)} 
\end{cases} ,
\end{eqnarray}
with boundary condition (\ref{modK}), on $ i = \pm (3 n _x +1) $ and $ j = \pm ( 3 n _y + 1 ) $
by the Gauss-Seidel iteration.  Once the Green's function, $ G ^\prime _{i,j,0,0} $, is given,
then the Green's function, $ G _{i,j,i^\prime,j^\prime} $, is obtained by summing up the
contributions from the mirror images.   Thus we can save the computation time for
solving the Poisson equation greatly by the method of mirror image.   The boundary condition
(\ref{modK}) improves the accuracy of the Green's function for a small wavenumber, $ kH \la 0.05 $.

When $ k = 0 $, Eq. (\ref{modK}) is replaced with
\begin{eqnarray}
G ^\prime _{i,j,i^\prime,j^\prime} & = & - 2 G \Delta x \Delta y \ln r ,  \label{modK0} 
\end{eqnarray}
where the asymptotic form of the modified Bessel function near $ z = 0 $ is applied. 
Although Equation (\ref{modK0}) contains an offset proportional to $ \ln (k/2) $,  
the $ x $- and $ y $-components of the gravity are not affected by the offset.

\section{Case of $ B _0 = 0 $  \label{1Dmode}}

When $ B _0 = 0 $, our equilibrium model is symmetric and  the perturbation equation is reduced 
to an ordinary differential equation if  we use the cylindrical coordinate, $ (r, \varphi, z) $.   Following Appendix C of paper I,
we express the density, displacement and potential in the form, 
\begin{eqnarray}
\rho & = & \rho _0 + \delta \varrho (r) \cos k z , \\
\mbox{\boldmath$\xi$} & = & \xi _r (r) \cos k z \mbox{\boldmath$e$} _r + \xi _z (r)
\sin k z \mbox{\boldmath$e$} _z , \\
\psi & = & \psi _0 + \delta \psi (r) \cos k z .
\end{eqnarray}
The perturbation equations are
written as
\begin{eqnarray}
\delta \varrho & = & - \frac{1}{r} \frac{\partial}{\partial r} \left( r \rho _0 \xi _r \right)
- k \rho _0 \xi _z , \label{noMrho} \\
\sigma ^2  \xi _r & = & - \frac{\partial}{\partial r} \left[ \left( \frac{dP}{d\rho} \right) \frac{\delta \varrho}{\rho _0} \right]
-   \frac{\partial}{\partial r} \delta \psi , \label{noMxir}  \\
\sigma ^2 \xi _z & = & k \left( \frac{\partial P}{\partial \rho} \right)  \frac{\delta \varrho}{\rho _0} + k  \delta \psi  , \label{noMxiz}  \\
4 \pi G \delta \varrho & = & \frac{1}{r} \frac{\partial}{\partial r} \left( r \frac{\partial}{\partial r} \delta \psi \right) 
- k ^2 \delta \psi . \label{noMPoisson}
\end{eqnarray}
From 
Equations (\ref{noMxir}) and (\ref{noMxiz}), we  obtain
\begin{eqnarray}
\xi _z & = & - \int  k \xi _r dr . \label{rotxi} 
\end{eqnarray}
We solve Equations (\ref{noMrho}), (\ref{noMxir}), (\ref{noMPoisson}), and (\ref{rotxi}) in the discretized form.

We express the perturbation using the radial displacement, 
$ \mbox{\boldmath$\xi$} _r = (\xi _{r,1/2}, \xi _{r, 3/2}, \dots, \xi _{r,n-1/2} )$, where $ \xi _{r,j-1/2} $ denotes
the radial displacement at $ r = (j-1/2) \Delta r $.  We will show that Equation (\ref{noMxir}) can
be expressed in the discretized form,
\begin{eqnarray}
\sigma ^2 \xi _{r,j-1/2} & = & \sum _{i} F _{ji} \xi _{r,i-1/2} , \label{radial}
\end{eqnarray}
where the matrix elements, $ F _{ji} $, are obtained by the following procedure.    
When evaluating $ F _{ji} $, we set
\begin{eqnarray}
\xi _{r,j-1/2} & = &
\begin{cases}
1 & \mbox{if}~j = i \\
0 & \mbox{otherwise}
\end{cases} .
\end{eqnarray}
Using equation (\ref{rotxi}) we obtain the longitudinal displacement,
\begin{eqnarray}
\xi _{z,j} & = & 
\begin{cases}
- k \Delta r & (j < i ) \\
0 & j \ge i
\end{cases} .
\end{eqnarray}
The change in the density is evaluated to be
\begin{eqnarray}
\delta \varrho _j & = & 
\begin{cases}
\displaystyle - \frac{4 \rho _{0,1/2} \xi _{r,1/2} }{\Delta r} - k \rho _{0,0} \xi_{z,0} & (j = 0) \\
\displaystyle - \frac{1}{r _j \Delta r} \left( r _{j+1/2} \rho _{0,j+1/2} \xi _{r,j+1/2} 
- r _{j-1/2} \rho _{0,j-1/2} \xi _{r,j-1/2} \right)
- k \rho _{0,j} \xi _{z,j} & (j \ne 0) 
\end{cases} ,
\end{eqnarray}
by discretizing Equation (\ref{noMrho}), where $ \delta \varrho _{j} $ and 
$ \rho _{0,j+1/2} $ denote the values at  {$ r = r _j $ and $ r _{j+1/2}$}, respectively. 
The change in the gravitational potential, $\delta \psi $, is obtained by solving 
the discretized Poisson equation
\begin{eqnarray}
4 \pi G \delta \varrho _{j} & = & 
\left\{ 
\begin{array}{ll}
\displaystyle 2 \frac{\delta \psi _1 - \delta \psi _0}{\Delta r ^2}
- k ^2 \delta \psi _0  & \hskip 20pt ( j = 0) \\
\displaystyle - \frac{ r  _{j+1/2} \delta \psi _{j+1} - 2 r _j 
\delta \psi _j + r _{j-1/2} \delta \psi _{j-1}}{r _j \Delta r ^2}
- k ^2 \delta \psi _j  & \hskip 20pt (j=1, 2, \dots , n)
\end{array}  \right. ,
\end{eqnarray}
with the boundary condition
\begin{eqnarray}
\delta \psi _{n+1} = \frac{K _0 [k (n+1) \Delta r]}{K _0 (k n \Delta r)} \delta \psi _{n} , 
\label{psi-boundary}
\end{eqnarray}
where $ K _0 $ denotes the modified Bessel function (see Appendix \ref{BesselK}).
By discretizing Equation (\ref{noMxir}) we obtain
\begin{eqnarray}
\sigma ^2 \xi _{r,j-1/2} & = & - \frac{1}{\Delta r} \left[ \left( \frac{dP}{d\rho} \right)_{j} 
\frac{\delta \varrho _{j}}{\rho _{0,j}} -  \left( \frac{dP}{d\rho} \right)_{j-1} 
\frac{\delta \varrho _{j-1}}{\rho _{0,j-1}} + \delta \psi _j - \delta \psi _{j-1} \right] .  \label{noMxird} 
\end{eqnarray}
The righthand side of Equation (\ref{noMxird}) denotes the matrix element, $ F _{ji} $.

When $ k = 0 $, we replace Equation (\ref{noMxird}) with
\begin{eqnarray}
\sigma ^2 \xi _{r,j-1/2} & = &
\displaystyle  - \frac{1}{\Delta r} \left[ \left( \frac{dP}{d\rho} \right)_{j} 
\frac{\delta \varrho _{j}}{\rho _{0,j}} -  \left( \frac{dP}{d\rho} \right)_{j-1} 
\frac{\delta \varrho _{j-1}}{\rho _{0,j-1}} \right] + 4 \pi G \rho _{0,j} \delta _{i,j}
\end{eqnarray}
where $ \delta _{i,j} $ denotes the Kronecker's delta, since
\begin{eqnarray}
\frac{\partial \psi}{\partial r} & = & - 4 \pi G \rho _0 \xi _r 
\end{eqnarray}
for $ k = 0 $.   

The growth rate, $ \sigma $, is obtained as the eigenvalue of the matrix, $ F _{ji} $.
The spatial resolution and outer boundary are set to be $ \Delta r = 0.1 H $ and
$ r _n \ge 60 H $.   The outer boundary should be set very far $ (r \ge 100 H) $
for $ p < 1 $.   Otherwise the growth rate is underestimated.

%\section{Variational Principle}




\clearpage

\allauthors

\listofchanges

\end{document}